\documentclass[a4paper,fleqn,usenatbib]{mnras}
\usepackage{rotating}
\usepackage{amsmath} 
\usepackage{amssymb}
\usepackage{xcolor}
\usepackage{hyperref}
\usepackage{tabularx}
\usepackage{graphicx} 
\usepackage{threeparttable}  
\usepackage{xfrac} 
\usepackage{CJKutf8} 
\usepackage{lineno}
\usepackage{multirow} 

\def\ltsima{$\; \buildrel < \over \sim \;$}
\def\simlt{\lower.5ex\hbox{\ltsima}}
\def\gtsima{$\; \buildrel > \over \sim \;$}
\def\simgt{\lower.5ex\hbox{\gtsima}}
\def\kpc{{\rm\,kpc}}

\def\pc{{\rm\,pc}}

\def\deg{^\circ}

\def\s{\ifmmode \widetilde \else \~\fi}
\def\={\overline}

\def\spose#1{\hbox to 0pt{#1\hss}}

\def\lta{\mathrel{\spose{\lower 3pt\hbox{$\mathchar"218$}}
     \raise 2.0pt\hbox{$\mathchar"13C$}}}
\def\gta{\mathrel{\spose{\lower 3pt\hbox{$\mathchar"218$}}
     \raise 2.0pt\hbox{$\mathchar"13E$}}}
\def\Dt{\spose{\raise 1.5ex\hbox{\hskip3pt$\mathchar"201$}}}    
\def\dt{\spose{\raise 1.0ex\hbox{\hskip2pt$\mathchar"201$}}}    

\def\dotsfill{\leaders\hbox to 1em{\hss.\hss}\hfill}

\title[]{Is M31 at the center of its satellite system ? }

\author[A. Doliva-Dolinsky et al. ]{Amandine Doliva-Dolinsky$^{1}$\thanks{E-mail: amandinedolinsky@gmail.com}, N. F. Martin$^{2,3}$, Michelle L. M. Collins$^{1}$ \\
$^1$University of Surrey, School of Mathematics and Physics, Guildford, GU2 7XH, UK \\
$^2$Universit\'e de Strasbourg, CNRS, Observatoire astronomique de Strasbourg, UMR 7550, F-67000 Strasbourg, France\\
$^3$Max-Planck-Institut f\"{u}r Astronomie, K\"{o}nigstuhl 17, D-69117 Heidelberg, Germany \\
}

\date{Accepted XXX. Received YYY; in original form ZZZ}

\pubyear{2023}

\begin{document}
\label{firstpage}
\pagerange{\pageref{firstpage}--\pageref{lastpage}}
\maketitle

\begin{abstract}
 The arrangement of M31's dwarf galaxies exhibits anisotropy, with the majority located in the hemisphere between the Milky Way and the host galaxy. This study aims to investigate whether M31’s present location is aligned with the center of its distribution of dwarf galaxies. We use forward modeling to infer the center of the M31 satellite 3D spatial distribution, folding in the completeness of dwarf galaxy searches. We observe a displacement of the center of the satellite distribution, relative to the center of M31, of approximately 10--50 kpc towards the Milky Way. Nonetheless, the center of M31 remains compatible with the center of the dwarf galaxy distribution given the broad constraints on its position, with the significance of the shift ranging from $\leq 1\sigma$ to $1.9\sigma$, depending on the assumed form of the volumetric spatial distribution of satellites. If M31 is truly offset from its satellite system, a quadrupling of the number of known satellites would be necessary to infer a significant ($3\sigma$) offset. Hence, expanding the number of known dwarf galaxies is crucial to deepen our understanding of the distribution of M31 satellites and further shed on its peculiar structure.
\end{abstract}


\section{Introduction}

\quad The distribution of M31 dwarf galaxies has been reported to be highly anisotropic, with a majority of the satellites situated closer to the MW rather than on the opposite hemisphere \citep{McConnachie2005b,Conn2012,Wan2020}. This peculiar feature has been reenforced by the updated satellite distances derived from RR-Lyrae in \cite{Savino2022}.  Although the detection of dwarf galaxies located farther away from the Milky Way poses a greater challenge, our previous work convincingly dismisses detection limits as the only explanation for this distribution, with a confidence level exceeding 99$\%$ \citep{Doliva2023}. Consequently, the observed anisotropy in the M31 satellite system is indeed a genuine feature. 

\quad Recently, \cite{Kanehisa2025} investigated the occurrence of such anisotropy in cosmological and galaxy evolution simulations, finding it to be very rare. While the implications of this distribution on cosmology, as well as the formation and evolution of M31, remain uncertain, one proposed explanation suggests that it could be attributed to the recent accretion of satellites enough time has not yet passed for the recently accreted to become phase mixed \citep{McConnachie2005}. Another plausible explanation arises from the anticipated displacement of approximately 15-25 kpc of the Milky Way from the center of its dark-matter halo \citep{Garavito-Camargo2021}, which can be attributed to the presence of the Large Magellanic Cloud (LMC). Considering the presence of the Giant Stream in the M31 halo \citep{Ibata2001, Ibata2014} and given that most M31 models predict a significant merger event has occurred $\sim2$ Gyr ago \citep{Hammer2018,DSouza2018}, this raises the question of whether M31 could be displaced from the center of its dark matter halo and satellite system and `sloshing' in its halo. Additionally, using 6D phase space data, \cite{Patel2025} compute the M33–M31 orbits and find that M33’s gravitational influence causes a displacement of M31’s center of mass by 30–150 kpc. Furthermore, by integrating the orbits of ten M31 satellites backward in an M31+M33 potential and then forward without M33’s influence, \cite{Patel2025} show that M33 is unlikely to be the main cause of the anisotropy. However, it may still contribute, along with other dynamical effects.

\quad We consider a scenario where the satellite galaxies are phase-mixed, while M31 is displaced within its dark matter halo due to a past merger or the gravitational influence of M33. In this case, the satellite distribution should be isotropic and follow the dark matter profile. To test for a possible offset of M31 from the dark matter center, we construct a simple model with each satellite equally contributing to the inferred shift in M31's center that uses forward modeling to locate the center of the satellite distribution without assuming it matches M31’s position.

\quad In Section~\ref{sample_articel3}, we describe the satellite sample and its completeness. Section~\ref{method} details the model and its implementation to obtain the results presented in Section~\ref{results} and discussed in Section~\ref{conclu}.

\begin{table}
\begin{center}
  \caption{Sample of the dwarf galaxies present in the PAndAS survey.}
\begin{tabular}{llll} 
   \hline
  Name &$\alpha$(J2000) & $\delta$(J2000) & $D_\mathrm{MW}(kpc)$ \\
  \hline
   \multicolumn{4}{c}{Within the PAndAS footprint} \\
And~I 		&00$^\textrm{h}$45$^\textrm{m}$39.7$^\textrm{s}$ 	&+38$\deg$02$'$15$''$  	&775$_{-17}^{+19}$\\
And~II 		&01$^\textrm{h}$16$^\textrm{m}$26.8$^\textrm{s}$  	&+33$\deg$26$'$07$''$    &667$_{-15}^{+16}$	\\
And~III   		&00$^\textrm{h}$35$^\textrm{m}$30.9$^\textrm{s}$  	&+36$\deg$29$'$56$''$    	&721$_{-16}^{+17}$	\\
And~V     		&01$^\textrm{h}$10$^\textrm{m}$17.5$^\textrm{s}$  	&+47$\deg$37$'$42$''$    	&759$_{-20}^{+21}$	\\
And~IX    		&00$^\textrm{h}$52$^\textrm{m}$53.4$^\textrm{s}$  	&+43$\deg$11$'$57$''$    	&702$_{-20}^{+19}$	\\
And~X     		&01$^\textrm{h}$06$^\textrm{m}$35.4$^\textrm{s}$  	&+44$\deg$48$'$27$''$   	&630$_{-18}^{+18}$	\\
And~XI    		&00$^\textrm{h}$46$^\textrm{m}$19.7$^\textrm{s}$ 	&+33$\deg$48$'$10$''$    &751$_{-22}^{+23}$	\\
And~XII   		&00$^\textrm{h}$47$^\textrm{m}$28.3$^\textrm{s}$ 	&+34$\deg$22$'$38$''$    	&718$_{-26}^{+25}$	\\
And~XIII  		&00$^\textrm{h}$51$^\textrm{m}$51.0$^\textrm{s}$  	&+33$\deg$00$'$16$''$    	&821$_{-26}^{+28}$	\\
And~XIV   	&00$^\textrm{h}$51$^\textrm{m}$35.0$^\textrm{s}$  	&+29$\deg$41$'$23$''$   	&773$_{-21}^{+21}$	\\
And~XV    	&01$^\textrm{h}$14$^\textrm{m}$18.3$^\textrm{s}$  	&+38$\deg$07$'$11$''$	&746$_{-18}^{+17}$	\\
And~XVI   	&00$^\textrm{h}$59$^\textrm{m}$30.3$^\textrm{s}$  	&+32$\deg$22$'$34$''$	&517$_{-19}^{+18}$	\\
And~XVII  	&00$^\textrm{h}$37$^\textrm{m}$06.3$^\textrm{s}$  	&+44$\deg$19$'$23$''$	&757$_{-23}^{+24}$	\\
And~XIX   	&00$^\textrm{h}$19$^\textrm{m}$34.5$^\textrm{s}$  	&+35$\deg$02$'$41$''$	&813$_{-31}^{+31}$	\\
And~XX    	&00$^\textrm{h}$07$^\textrm{m}$30.6$^\textrm{s}$ 	&+35$\deg$07$'$37$''$	&741$_{-27}^{+27}$	\\
And~XXI   	&23$^\textrm{h}$54$^\textrm{m}$47.9$^\textrm{s}$  	&+42$\deg$28$'$14$''$	&770$_{-22}^{+23}$	 \\
And~XXII  	&01$^\textrm{h}$27$^\textrm{m}$40.4$^\textrm{s}$  	&+28$\deg$05$'$25$''$	&754$_{-23}^{+24}$ \\
And~XXIII 	&01$^\textrm{h}$29$^\textrm{m}$21.0$^\textrm{s}$  	&+38$\deg$43$'$26$''$	&745$_{-25}^{+24}$	\\
And~XXIV 	&01$^\textrm{h}$18$^\textrm{m}$32.7$^\textrm{s}$  	&+46$\deg$22$'$13$''$	&609$_{-20}^{+19}$	\\
And~XXV  	&00$^\textrm{h}$30$^\textrm{m}$09.9$^\textrm{s}$  	&+46$\deg$51$'$41$''$	&752$_{-23}^{+23}$	\\
And~XXVI  	&00$^\textrm{h}$23$^\textrm{m}$46.3$^\textrm{s}$  	&+47$\deg$54$'$43$''$	&786$_{-23}^{+24}$	\\
And~XXX   	&00$^\textrm{h}$36$^\textrm{m}$34.6$^\textrm{s}$  	&+49$\deg$38$'$49$''$	&558$_{-16}^{+17}$	\\
NGC~147   	&00$^\textrm{h}$47$^\textrm{m}$27.0$^\textrm{s}$  &+34$\deg$22$'$29$''$	&773$_{-20}^{+21}$	\\
NGC~185   	&00$^\textrm{h}$38$^\textrm{m}$58.0$^\textrm{s}$  &+48$\deg$20$'$15$''$	&650$_{-18}^{+18}$	\\
\multicolumn{4}{c}{Outside the PAndAS footprint} \\
LGS~3/Pisces &01$^\textrm{h}$03$^\textrm{m}$55.0$^\textrm{s}$ 	&+21$\deg$53$'$06$''$	&605.3$\pm14$ \\
And~VI		&23$^\textrm{h}$51$^\textrm{m}$46.4$^\textrm{s}$ 	&+24$\deg$35$'$11$''$	&831.8$\pm$23\\
And~VII		&23$^\textrm{h}$26$^\textrm{m}$31.7$^\textrm{s}$ 	&+50$\deg$40$'$33$''$	&824.1$\pm$23\\
And~XXIX	&23$^\textrm{h}$58$^\textrm{m}$55.6$^\textrm{s}$ 	&+30$\deg$45$'$20$''$	&711.2$^{+20}_{-19}$\\
And~XXXI	&22$^\textrm{h}$58 $^\textrm{m}$6.3$^\textrm{s}$ 	&+41$\deg$17$'$28$''$	&744.7$\pm$17\\
And~XXXII	&00$^\textrm{h}$35$^\textrm{m}$59.4$^\textrm{s}$ 	&+51$\deg$33$'$35$''$	&801.7$^{+23}_{-22}$\\
 And~XXXIII	&03$^\textrm{h}$01$^\textrm{m}$23.6$^\textrm{s}$ 	&+40$\deg$59$'$18$''$	&704.7$^{+20}_{-19}$\\

  \end{tabular}
   \label{sample_article3}
  \end{center}
\textbf{Notes:} All distances are from \cite{Savino2022}. Some dwarf galaxies, although in the considered footprint, are not part of this sample because the detection limits for their survey and/or at their location are not quantified (M21, NGC205, Pisces VII, Pegasus V), because of significant uncertainties in their structural parameters (And~XXVII; \citealt{Richardson2011}), or because their distances from M31 exceed $300\kpc$ (And~XVIII; And~XXVIII; \citealt{Savino2022}).
\end{table}

\section{Sample}\label{sample_articel3}

\quad Our study is based on dwarf galaxies found within a projected radial distance of 30 kpc to a 3D radial distance of 300 kpc from M31. This range is determined based on an approximate estimate of M31's virial radius, with the inclusion of an empirical inner boundary that considers the difficulties associated with detecting dwarf galaxies near M31's disk \citep{Doliva2022}. Given the importance of the MW stellar contamination at higher latitude, north of M31, and its impact on the recovery of dwarf galaxies, we also impose a Galactic latitude cut $b<-9.5\deg$ \citep{Doliva2022}. The resulting list of dwarf galaxies is presented in Table~\ref{sample_article3}. Considering the uncertain nature of And XXVII \citep{Preston2019}, as well as the significant uncertainties in its structural properties \citep{Richardson2011, Martin2016}, we have decided to exclude it from our sample. The distances are derived from the RR Lyrae measurements and are taken from \cite{Savino2022}.

\quad The sample of dwarf galaxies is split into two categories: those that fall within the deeper PAndAS footprint and those located outside of it. For dwarf galaxies outside the PAndAS footprint, our selection is limited to dwarf galaxies that have been detected either in SDSS or in Pan-STARRS, because of the necessity to have an assessment of the search incompleteness. As a result, the recently discovered dwarf galaxies Pisces VII, Pegasus V, or Andromeda~XXXV \citep{Collins2022, Collins2023,Arias2025}, are not considered in this study. 

Inside the PAndAS survey, the completeness limits were determined by \cite{Doliva2022} through the injection of the stars of approximately half a million artificial dwarf galaxies into the PAndAS catalogue. The search algorithm developed by \cite{Martin2013} was used to locate spatial and photometric overdensities of old metal-poor stars, and determine the recovery (or lack thereof) of a dwarf galaxy. Specifically, for each of the $\sim400$ MegaCam fields in PAndAS, the recovery fraction was assessed for dwarf galaxies with $-8.5<M_V<-4.5$ and $1.8<\log(r_\mathrm{h}(\pc))<3.0$. To facilitate the calculation of the recovery fraction for any galaxy at a given location, magnitude, and size, an analytical model is fitted to the resulting $M_V-\log(r_\mathrm{h}(\pc))$ recovery fraction grid. Furthermore, an additional analytical model was developed to account for the influence of the heliocentric distance of a dwarf galaxy on recovery fractions. While the impact of the distance is less prominent compared to the size, magnitude, or location of the dwarf galaxy, it is still a factor that cannot be disregarded \citep{Doliva2022}.

\quad Given the comparable depth of the SDSS \citep{SDSS2003} and Pan-STARRS \citep{Pan-STARR12016} surveys, the recovery fractions for dwarf galaxies located outside the PAndAS footprint, are based on the detection limit derived by \cite{Koposov2008} using SDSS data. To assess the detectability, \cite{Koposov2008} inject artificial dwarf galaxies into the SDSS catalogue and their detection or non-detection is determined using a matched-filter algorithm. In this study, the completeness is evaluated across a range of $-11\leq M_V\leq0$, $0\leq\log(r_\mathrm{h}(\pc))\leq3.0$. Given the distance of M31, we exclusively use the results obtained for distances between 512 and 1024 kpc. An analytical model is fitted to the resulting $M_V-\log(r_\mathrm{h}(\pc))$ recovery fraction grid. We extrapolate both PAndAS and SDSS analytical model for detection limits to cover a range of $-17< M_V< -5.5$ and $1.8<\log(r_\mathrm{h}(\pc))<4.0$.

\section{Method} \label{method}

\begin{table*}
\begin{center}
  \caption{Values for the models parameters depending on the dwarf galaxy sample and the position prior}
\begin{tabular}{lllllll} 
   \hline
  Model & Prior on the & Sample & $\Delta$x (kpc) & $\Delta$y (kpc) & $\Delta$D (kpc) & $\gamma$ \\
   & center position &  & &  &  & \\
  \hline
  Power law 	& Uniform 	& PAndAS 		& 3$^{+10}_{-11}$ 			&8$\pm12$		&-46$_{-30}^{+35}$		&$-2.0\pm0.10$ \\
  Power law 	& Gaussian 	& PAndAS  		& 3$^{+8}_{-9}$ 			&5$\pm10$		&-14$^{+21}_{-24}$		&$-2.0\pm0.10$ \\
  Power law 	& Gaussian 	& PAndAS + SDSS 	& 2$^{+7}_{-8}$ 			&4.5$\pm10$		&-18$^{+22}_{-27}$		&$-2.0\pm0.10$ \\
  NFW 		& Gaussian 	& PAndAS 		& 4$^{+3}_{-3}$ 			&1$\pm4$			&-27$\pm14$			& \\
\end{tabular}
\label{param_article3}
\end{center}
\end{table*}

\subsection{Basics of the model}
\quad We aim to infer the 3-dimensional center of the spatial distribution of M31 satellites while accounting for the dwarf galaxy search completeness. We consider a sample of $N$ dwarf galaxies, listed in Table~\ref{sample_article3}, defined by the set of properties properties $\mathcal{D}=\{\overrightarrow{d_k}\}_{1\leq k\leq N}$. The probability, $P_{sp}$, that those satellites follow a 3D spatial density distribution model, $\rho_{sp}$, defined by the set of parameters $\mathcal{P}$ over a region $\mathcal{A}$ is therefore 

\begin{eqnarray}
\label{likelihoodsingle}
P_{sp}\left(\mathcal{D}|\mathcal{P}\right)&=&\prod_k P_{sp,k}\left(\overrightarrow{d_k}|\mathcal{P}\right),\\
&=&\frac{\rho_{sp}\left(\overrightarrow{d_k}|\mathcal{P}\right)}{\int_\mathcal{A}\rho_{sp}\left(\overrightarrow{d_k}|\mathcal{P}\right) \tau\left(\overrightarrow{d_k}\right) d\mathcal{\overrightarrow{d_k}}}, \nonumber
\end{eqnarray}
with $\tau\left(\overrightarrow{d_k}\right)$ a completeness coefficient taking into account the footprint of the survey and the surface brightness detection limits. This factor is obtained by marginalizing the recovery fraction of dwarf galaxies over their magnitude and size. We use the recovery fraction from \cite{Doliva2022} within the PAndAS footprint, and from \cite{Koposov2008} for the rest of the sky. The marginalization is performed using the luminosity function and the size–luminosity relation from \cite{Doliva2023}, marginalizing over their uncertainties.

\quad Following the principle of Bayesian inference, first described by \cite{Bayes1753} and further developed by \cite{Laplace1812}, the likelihood we aim to determine, $P\left( \mathcal{P}|\mathcal{D}\right)$, is link to $P\left( \mathcal{D}|\mathcal{P} \right)$ via the prior $P(\mathcal{P})$ such that
\begin{equation}\label{Bayes}
P(\mathcal{P}|\mathcal{D}) \propto P\left( \mathcal{D}|\mathcal{P} \right)P(\mathcal{P}).
\end{equation}

\subsection{Volumetric Spatial Distribution}

\quad Throughout this article, we assume an isotropic distribution of satellites, and we choose to infer the center of the distribution for two models. The dwarf galaxies are defined by their sky coordinates ($\alpha_{gal}$, $\delta_{gal}$) and their heliocentric distance $D_{MW,gal}$, therefore $\overrightarrow{d_k}=\{\alpha_{gal,k},\delta_{gal,k}, D_{MW,gal,k}\}_{1\leq k\leq N}$. As all models tested are isotropic, a simplification can be made by defining the dwarf galaxies by $r_c$, their distance to the center of the distribution, yielding $\overrightarrow{d_k^{\prime}}=\{r_{c,k}\}_{1\leq k\leq N}$. Those values are obtained through simple algebra by using the coordinates of the center of the dwarf galaxy (i.e, its sky coordinates ($\alpha_c$, $\delta_c$) and its heliocentric distance $D_{MW,c}$).

\quad First, we choose to test an agnostic and simple spatial distribution function. The density is assumed to follow a power law as a function of the radius $r_c$. As the distribution is isotropic, we have
\begin{eqnarray}
\rho(r_c|\gamma_{pow})&=&\int_0^{2\pi}\int_0^{\pi} r_c^2 r_c^{\gamma_{pow}}\sin(\theta)d\theta d\phi.\\
&=&4\pi r_c^{2+\gamma_{pow}} \nonumber
\end{eqnarray}

\quad Secondly, we assume an NFW distribution function \citep{Navarro1996} as it is expected that satellites follow the NFW profile of the dark matter \citep{Guo2013}. The virial radius $R_{vir}$ is roughly estimated at 300 kpc and the mass of M31, $M_{M31}$, is chosen to be $1.33\times10^{12}M_\odot$ \citep{Penarrubia2016}, yielding a concentration $c\sim7.9$ following the relation presented in \cite{Diemer2019}. The density can be written 
\begin{eqnarray}
\rho(r_c)&=&\frac{\rho_0}{\frac{r_c}{R_s}\left(1+\frac{r_c}{R_s}\right)^2}, \\
\mathrm{with}\; \rho_0&=&\frac{M_{M31}}{4\pi R_s^3\left(\ln(1+c)-\frac{c}{1+c}\right)} \;\mathrm{and}\; R_s=\frac{R_{vir}}{c}. \nonumber
\end{eqnarray}

 yielding, 
\begin{eqnarray}
\mathcal{P}_{pow}&=&\{\gamma_{pow}, \alpha_c, \delta_c, D_{MW,c} \},\\
\mathcal{P}_{NFW}&=&\{\alpha_c, \delta_c, D_{MW,c} \}.
\end{eqnarray}

\subsection {Priors}
\quad Following Equation~\ref{Bayes}, we set a prior for each parameter to determine $P(\mathcal{P}|\mathcal{D})$. 
\begin{itemize}
\item We impose $\gamma_{pow}\leq-2$ as it is expected that the density of dwarf galaxies decreases as a function of the distance to the main host. We therefore choose to impose a uniform prior on $\gamma_{pow}$ such that $-10\leq \gamma_{pow} \leq -2$. 
\item \cite{Salomon2023} present a prediction regarding the maximum/average distance between the center of mass of the disk and the center of mass of the DM halo/satellite distribution, estimating it to be 30 kpc. Based on this, we choose to impose Gaussian priors on the physical coordinates of the center in the tangential plane to M31, denoted as $x_c$ and $y_c$. These Gaussian priors have a mean of 0 and a dispersion, $\sigma_{offset}$, equal to $\frac{30}{\sqrt{3}}$ kpc in each of the 3 spatial directions. These priors can be easily converted into priors on $\alpha_c$ and $\delta_c$ via the distance of M31.
\item For the distance of the distribution center to the MW, $D_{MW,c}$, we choose to account for the possible offset with the center of M31 along with the uncertainty on M31's distance. We therefore choose a Gaussian prior with a mean of 0 and a dispersion $\sigma$ such that $\sigma=\sqrt{\sigma_{offset}^2+\sigma_{D_{M31}}^2}$. As determined by \cite{Savino2022}, $\sigma_{D_{M31}}$=22 kpc and as detailed above $\sigma_{offset}=\frac{30}{\sqrt{3}}$.

\end{itemize}

\subsection{Implementation}

\begin{figure*}
\begin{center}
\includegraphics[width=0.8\hsize]{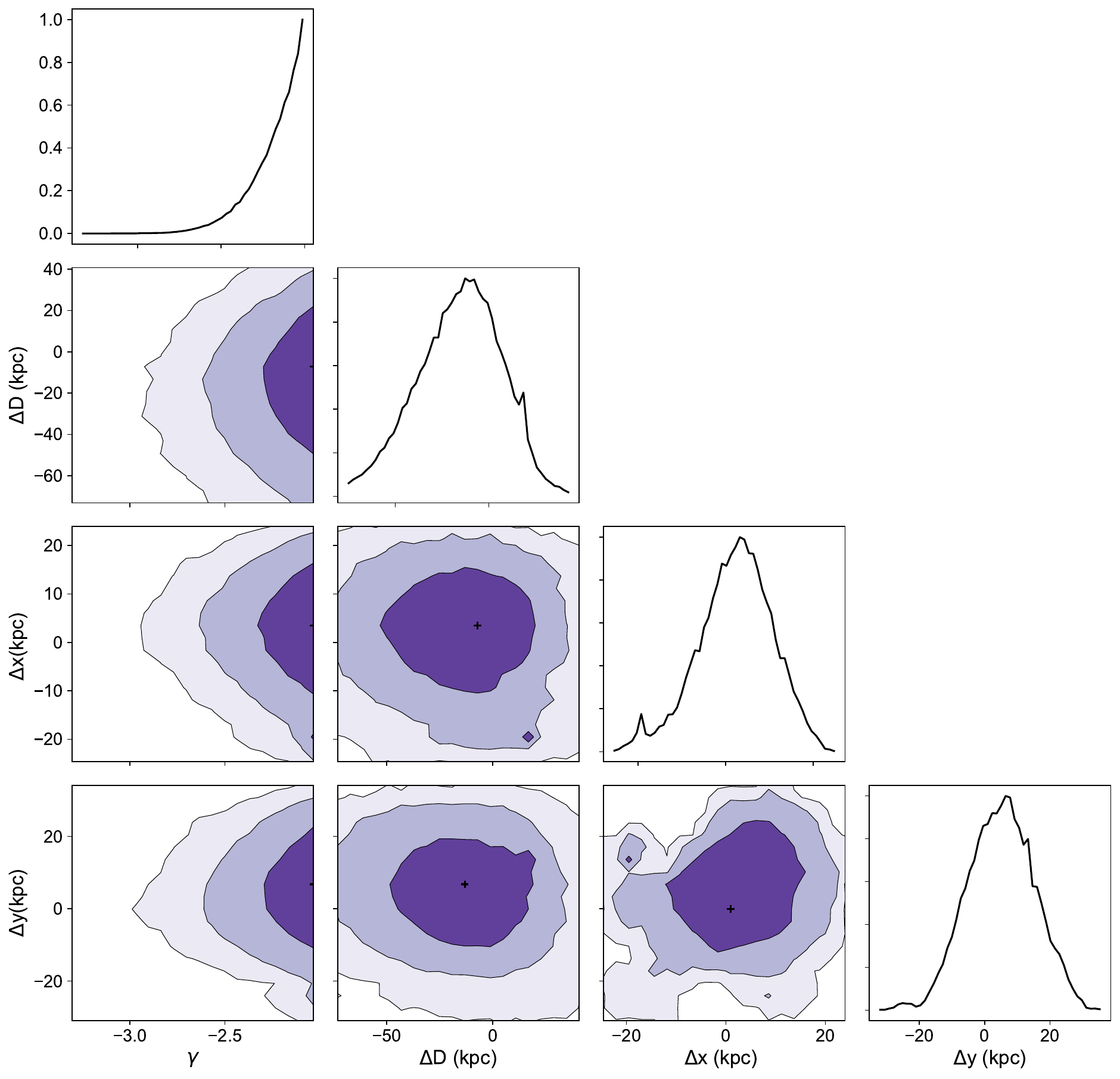}
\caption{\label{corner_article2} \textit{Left panels}: Marginalized PDFs and correlation graphs for each parameter of the model composed by a power-law spatial distribution when enforcing a Gaussian prior on the central position parameters and when using the PAndAS sample. }
\end{center}
\end{figure*}

\begin{figure*}
\begin{center}
\includegraphics[width=1\hsize]{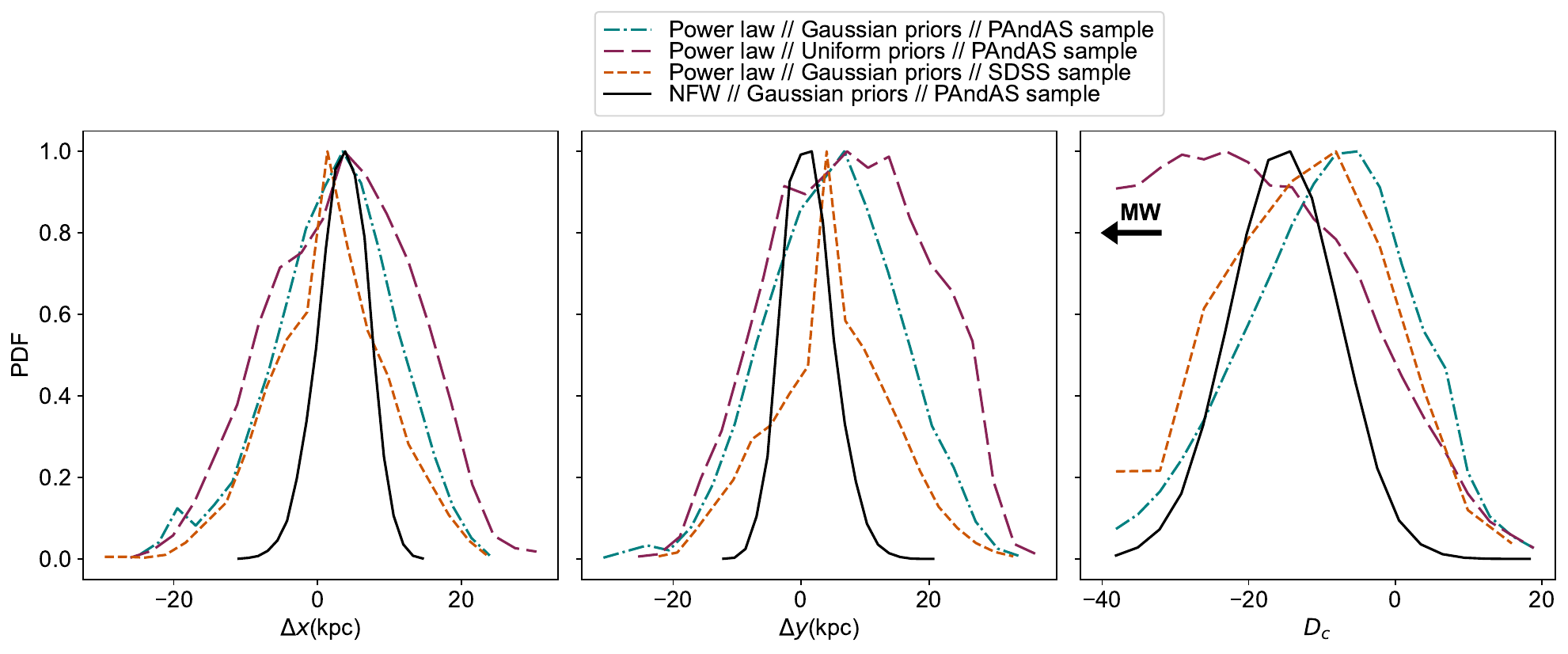}
\caption{\label{corner_article3}Marginalized PDFs of the distance between the center of M31 and the one of the satellites spatial distribution. While the favored distance of the center are shifted toward the MW, the center of the distribution is overall compatible with M31 center.}
\end{center}
\end{figure*}

\begin{figure*}
\begin{center}
\includegraphics[width=0.9\hsize]{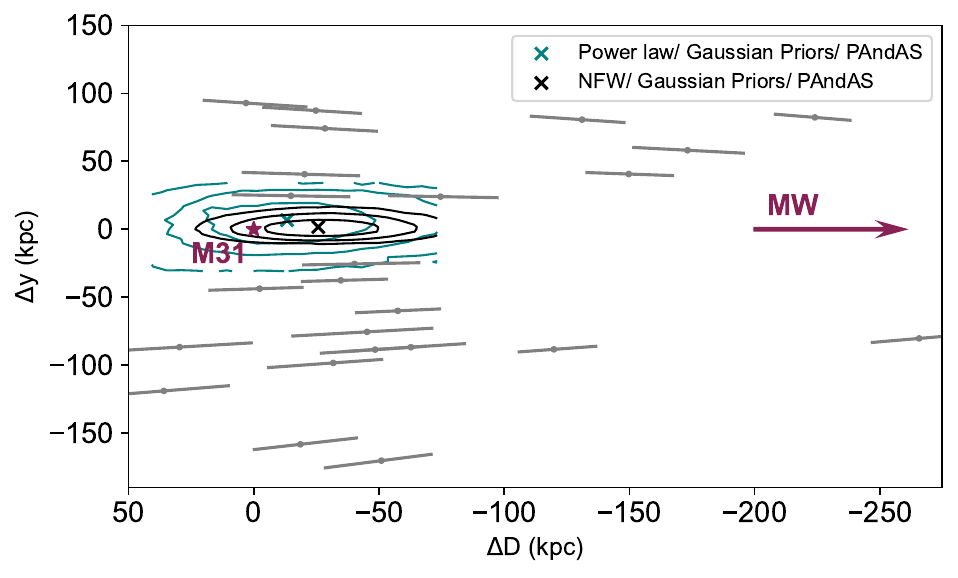}
\caption{\label{obsdensity}  PDFs for the center's position in the y-z plane in the case of a power law and NFW spatial distribution. M31's position is marked by a star, known dwarf galaxies are represented by purple dots, and the inferred satellite distribution center positions are denoted by blue/black crosses, along with their associated confidence contours (68\%, 90\%, 95\%) . We infer that M31 center is compatible with the center of the distribution but the model favors a center shifted toward the MW. }
\end{center}
\end{figure*}

\quad The likelihood is sampled using our own Metropolis-Hasting algorithm \citep{Metropolis1953,Hastings1970}. To accurately account for the uncertainties associated with the observed properties of the dwarf galaxies, we incorporate the probability distribution functions (PDFs) for each parameter instead of relying on single values. Following \cite{Conn2013}, we convolve the likelihood for a single value expressed in Equation~\ref{likelihoodsingle} with the PDFs of distances of M31 satellites to the MW such that 
\begin{eqnarray}
\mathcal{L}(\mathcal{D}^\prime_\Omega|\mathcal{P}^\prime)=\int_\Omega \mathcal{L}(\mathcal{D}^\prime|\mathcal{P}^\prime)g(\mathcal{D}^\prime)d\mathcal{D}^\prime,
\end{eqnarray}
with $\Omega$ being the set of all combinations of distances $\mathcal{D^\prime}$, and $g(\mathcal{D}^\prime)$ the probability of a given combination. To compute this integral, we employ a Monte-Carlo method. We generate 50 satellite systems by drawing from the PDF of $D_{MW,gal}$ for each galaxy in our sample. The final distribution is obtained by summing the resulting chains.

\section{Results} \label{results}

\quad For all spatial distribution models, Table~\ref{param_article3} provides constraints on the center position and slope, where applicable. The equatorial coordinates of the distribution center are projected onto the tangential plane to the centre of M31 in order to present the parameters as the shift of the distribution center from M31. We refer to these sky coordinates as ($\Delta x$, $\Delta y$), and the shift in distance along the line of sight with respect to the MW as $\Delta D$. A negative $\Delta D$ corresponds to a shift of the center of the distribution toward the MW. To determine the favored value, we consider the peak of the marginalized PDF, while the credible interval are bound by the parameter values whose PDF values are 0.61 of the maximum. However, to use these results effectively, we strongly recommend using the MCMC chains available at \url{https://github.com/dolivadolinsky}. 

\quad Figure~\ref{corner_article2} shows the correlation between parameters and the marginalized PDF for each parameter in the case of a power-law spatial distribution, incorporating a Gaussian prior on the position and using the PAndAS sample, and Figure~\ref{corner_article3} presents the marginalized PDFs of the positional parameters of the center for each of the models tested. Although the three positional parameters are generally well constrained, the posteriors are rarely perfect Gaussians due to the non-trivial impact of the detection limits on the model. 

\quad In all cases with a power-law spatial distribution, we infer a favored slope of $\gamma=-2.0\pm0.1$. Although this agreement might suggest compatibility between the cases, since the favored slope is the lower bound of our uniform prior, it actually indicates that the limited number of dwarf galaxies is insufficient to adequately constrain this parameter.

\quad We infer a marginal shift in sky coordinates ranging from $\Delta x=2^{+7}_{-8}$ kpc to $\Delta x=4^{+3}_{-3}$ kpc and from $\Delta y=1\pm4$ kpc to $\Delta y=8\pm12$ kpc. 

\quad When using a power-law spatial distribution, a uniform prior on the position of the center, and the PAndAS sample, we infer a shift in the heliocentric distance of $\Delta D=-46^{+35}_{-30}$ kpc. However, the distance difference decreases to $\Delta D=-14^{+21}_{-24}$ kpc when a Gaussian prior is imposed, considering the PAndAS sample alone, and to $\Delta D=-18^{+22}_{-27}$ kpc when both the PAndAS and SDSS samples are used. This increase in distance suggests that the available data are not robust enough to override our prior preference for the center to be in proximity to M31, or that the center of the satellites located in the PAndAS footprint (and therefore closer in projection) is more affected by the anisotropy than the distant ones present in the SDSS/Pan-STARRS. However, when employing an NFW profile, which removes one degree of freedom, we infer a shift in the distance of $\Delta D=-27\pm14$ kpc.

\section{Discussion} \label{conclu}

\quad In this paper, we use forward modeling to infer the center and, if applicable, the slope of the spatial distribution of M31 satellites, assuming an isotropic distribution. Our analysis incorporated the detection limits of the PAndAS and SDSS surveys. 

\quad We conclude that the sky coordinates of the distribution center ($\alpha_c$, $\delta_c$) are consistent with the position of M31. This suggests that any potential offset of M31 from its satellite system would primarily be along the axis pointing towards the MW. We infer that the shift in the distance to the center of the satellite system compared to M31 is ranging between $\Delta D=-46^{+35}_{-30}$ to $\Delta D=-14^{+21}_{-24}$ kpc. While the favored value for M31 \citep{Savino2022} is compatible in the case of a power-law distribution with the Gaussian prior, it tentatively falls outside the $1\sigma$ range when using the NFW profile (Figure~\ref{obsdensity}), with an inferred shift toward the MW that is significant at the $1.9\sigma$ level. To establish a more robust constraint on the center and profile of this satellite distribution, it is crucial to increase the number of known dwarf galaxies, particularly beyond the coverage of the PAndAS footprint. Everything else remaining equal, a quadrupling of the number of satellites would be necessary for the offset of $\Delta D=-46^{+35}_{-30}$ kpc to become significant at the $3\sigma$ level (shrinking the uncertainties by 2). Expanding the sample will provide a deeper understanding of this peculiar satellite distribution and enable more comprehensive insights on galaxy formation and on the cosmology.

\quad We note that the isotropic satellite distribution used here is a simplification, and that more realistic, possibly anisotropic models may be needed—especially if the satellites are not phase-mixed. Moreover, M31’s mass estimates vary widely \citep{Bhattacharya2023}, we selected a value at the low end but choosing an higher M31 mass could influence the inferred shift. Nonetheless, these factors do not significantly affect our principal conclusion, which is limited by the scarcity of data to explore the anisotropy thoroughly. Our toy model does not capture the full satellite dynamics and consider that each satellite equally contributes to the inferred shift in M31's center, largely because only a few proper motion measurements exist for Andromeda’s dwarfs \cite{Sohn2020,Brunthaler2005,Brunthaler2007}. Incorporating such dynamical data will be vital to resolve the causes of the anisotropy.

\quad We cannot confidently conclude that M31 is not at the center of its satellite distribution in any of the cases studied. The previous findings reported in \cite{Doliva2023} indicated a rejection at a $99.9\%$ confidence level of completeness as the sole explanation for the observed anisotropy, showing that it is a real feature of the satellite system. While these results may appear contradictory, it is important to note the differing approaches employed in the two contributions. In the previous study, the center of the distribution was fixed to the position of M31, allowing for the inference of the distribution's slope and subsequent derivation of the probability that our inferred model exhibits the same level of anisotropy as M31. Conversely, our current study focuses on investigating both the position of the center and, if applicable, the slope of the distribution. We find that, considering the limited number of known dwarf galaxies, the center of M31 remains compatible with the center of its satellite distribution. However, it is worth noting that the preferred center position still exhibits a slight shift towards the Milky Way. To conclude, as the models and questions are different, it is not unexpected that the conclusion differ slightly. 

\quad Finally, the inferred distance results in a shift of approximately 10 to 50 kpc between M31 and the center of the satellite distribution. The magnitude of this shift is comparable to the effect expected to be caused by the Large Magellanic Cloud on the Milky Way, where the outer halo ($>30$ kpc) shifts from the center of mass (COM) of the disk by approximately 15 to 25 kpc \citep{Garavito-Camargo2021}. Similarly, the interaction of M31 with the progenitor of what is now the Giant stream, as described by \cite{Hammer2018} and \cite{DSouza2018}, may have contributed to the observed shift between M31 and the center of the satellite distribution. 

\section*{Acknowledgments}

 MLMC acknowledges support from STFC grants ST/Y002857/1 and ST/Y002865/1.
 ADD acknowledges support from STFC grants ST/Y002857/1.

\section*{Data Availability}
The data underlying this article are available in the article and in its online supplementary material at \url{https://github.com/dolivadolinsky}.

\bibliography{biblio.bib}
\bibliographystyle{mnras}
\end{document}